\pdfoutput=1

\documentclass[11pt]{article}

\usepackage[final]{acl}

\usepackage{times}
\usepackage{latexsym}
\usepackage[T1]{fontenc}
\usepackage[utf8]{inputenc}
\usepackage{microtype}
\usepackage{inconsolata}
\usepackage{graphicx}

\usepackage[normalem]{ulem}
\usepackage{booktabs}
\usepackage{multirow}
\usepackage{array}
\usepackage{rotating}
\usepackage{makecell}
\usepackage{vcell}
\usepackage{enumitem}
\usepackage{amssymb}

\newcolumntype{M}[1]{>{\raggedright\arraybackslash}m{#1}}

\usepackage{float}
\makeatletter
\renewcommand{\paragraph}{%
  \@startsection{paragraph}{4}%
  {\z@}{0.7ex \@plus .2ex \@minus .2ex}{-1em}%
  {\normalfont\normalsize\bfseries}%
}
\makeatother

\title{Understanding and Meeting Practitioner Needs When Measuring \\ Representational Harms Caused by LLM-Based Systems}

\author{Emma Harvey\thanks{Work conducted during a Microsoft Research internship.}$^{\diamondsuit}$ \quad Emily Sheng$^{\boxplus}$ \quad Su Lin Blodgett$^{\boxplus}$ \quad Alexandra Chouldechova$^{\boxplus}$
\\
\textbf{Jean Garcia-Gathright}$^{\boxplus}$ \quad \textbf{Alexandra Olteanu}$^{\boxplus}$ \quad \textbf{Hanna Wallach}$^{\boxplus}$
\\ $^{\diamondsuit}$Cornell University \quad $^{\boxplus}$Microsoft Research \\
\texttt{evh29@cornell.edu} \\
\texttt{\{emilysheng, sulin.blodgett, alexandrac,}
\\
\texttt{jeang, alexandra.olteanu, wallach\}@microsoft.com}}

\begin{document}
\maketitle
\begin{abstract}
The NLP research community has made publicly available numerous instruments for measuring representational harms caused by large language model (LLM)-based systems. These instruments have taken the form of datasets, metrics, tools, and more. In this paper, we examine the extent to which such instruments meet the needs of practitioners tasked with evaluating LLM-based systems. Via semi-structured interviews with 12 such practitioners, we find that practitioners are often unable to use publicly available instruments for measuring representational harms. We identify two types of challenges. In some cases, instruments are \textit{not useful} because they do not meaningfully measure what practitioners seek to measure or are otherwise misaligned with practitioner needs. In other cases, instruments---even useful instruments---are \textit{not used} by practitioners due to practical and institutional barriers impeding their uptake. Drawing on measurement theory and pragmatic measurement, we provide recommendations for addressing these challenges to better meet practitioner needs.\looseness=-1
\end{abstract}

\section{Introduction}\label{s-introduction}
\label{sec:introduction}
Representational harms~\citep{barocas2017, crawford2017trouble} occur when a system ``represents some social groups in a less favorable light than others, demeans them, or fails to recognize their existence altogether''~\cite{blodgett-etal-2020-language}.  Numerous studies have documented representational harms caused by large language model (LLM)-based systems~\citep[e.g.,][]{sheng-etal-2019-woman, dev-etal-2021-harms, venkit-etal-2022-study, kotek_2023_gender, hofmann_dialect_2024}. It is important to measure and mitigate such harms---especially for systems that will be deployed in real-world contexts. However, this is known to be a challenging task. Like many other concepts related to the capabilities, behaviors, and impacts of LLM-based systems, representational harms are abstract and can have contested meanings across use cases, languages, and cultures~\cite{wallach2025positionevaluatinggenerativeai}. As a result, they are particularly difficult to define precisely and thus measure~\cite{blodgett-etal-2020-language, dev-etal-2022-measures, katzman_2023_taxonomizing, wang2023measuring}.\looseness=-1

To facilitate measuring representational harms, the NLP research community has produced numerous publicly available\footnote{By \textit{publicly available}, we mean available on the internet or via an academic publication for others to use or adapt.} measurement instruments, including datasets, metrics, tools, benchmarks,\footnote{\textit{Benchmarks} consist of datasets and metrics.} and annotation instructions. In this paper, we investigate whether these instruments meet the needs of practitioners tasked with evaluating LLM-based systems. As studies examining the uptake of other responsible AI artifacts have found, practitioner needs as assumed in the research literature are often different from those actually voiced by practitioners~\cite{holstein_improving_2019, lee_landscape_2021, deng_exploring_2022, ojewale_towards_2024}. This potential mismatch presents a critical opportunity for researchers and practitioners to engage with one another. If measurement instruments do not meet the needs of practitioners tasked with evaluating LLM-based systems, \textit{those instruments will not be used by such practitioners}---causing developers and deployers of LLM-based systems either to not engage in measurement or to rely on bespoke instruments that might not be publicly understood.\looseness=-1

\begin{table*}[!t]
\begin{small}
\centering
\def\arraystretch{0.9}
\setlength{\tabcolsep}{0.3em}
\begin{tabular}{
>{\hspace{0pt}}p{0.1\linewidth}
>{\hspace{0pt}}p{0.87\linewidth}@{}} 

\toprule

\textbf{Instrument} & \textbf{Examples} \\

\midrule

Datasets & Fifty Shades of Bias~\cite{hada-etal-2023-fifty}, FairPrism~\cite{fleisig-etal-2023-fairprism}, ToxiGen~\cite{hartvigsen-etal-2022-toxigen}\\
\midrule
Metrics & WEAT~\cite{caliskan_semantics_2017}, SEAT~\cite{may-etal-2019-measuring}, $\alpha$-Intersectional Fairness~\cite{maheshwari-etal-2023-fair}
\\
\midrule
Tools
& Perspective API~\cite{lees_new_2022}, Llama Guard~\cite{inan2023llamaguardllmbasedinputoutput}, HateBERT~\cite{caselli-etal-2021-hatebert} \\ 
\midrule
Benchmarks & BOLD~\cite{dhamala_bold_2021}, BBQ~\cite{parrish-etal-2022-bbq}, StereoSet~\cite{nadeem-etal-2021-stereoset}\looseness=-1
\\
\midrule
Annotation\newline instructions & Included as part of instruments like datasets \cite[e.g.,][]{fleisig-etal-2023-fairprism} and benchmarks \cite[e.g.,][]{nadeem-etal-2021-stereoset}, or released as part of measurement frameworks \cite[e.g.,][]{magooda_framework_2023} \\
\midrule
Other & Matched guise probing~\cite{hofmann_dialect_2024}, DivDist~\cite{bommasani_trustworthy_2022} \\
\bottomrule
\end{tabular}
\caption{Examples of publicly available instruments for measuring representational harms.}
\label{t-instruments}
\vspace{-0.3cm}
\end{small}
\end{table*}

Through a series of semi-structured interviews with 12 practitioners tasked with evaluating LLM-based systems for representational harms, we find that practitioners are often unable to use publicly available measurement instruments, despite a desire to do so. We identify two types of challenges that lead to this. In some cases, instruments are \textit{not useful}: they do not meaningfully measure what practitioners seek to measure or are otherwise misaligned with practitioner needs. In other cases, instruments---even useful instruments---are \textit{not used} by practitioners due to practical or institutional barriers impeding their uptake. Although we focus on instruments for measuring representational harms, many of our findings apply broadly to cases where practitioners seek to use publicly available instruments to measure other abstract or contested concepts. However, our targeted focus allows us to identify cases where challenges are exacerbated by the specifics of measuring representational~harms~or evaluating LLM-based systems.\looseness=-1

Developing measurement instruments that are simultaneously useful and used is not a new challenge. Measurement theory from the social sciences has long been concerned with designing instruments that are useful, i.e., those that meaningfully measure what they purport to measure~\cite{adcock_measurement_2001, jacobs_measurement_2021, wallach2025positionevaluatinggenerativeai}. Pragmatic measurement builds on measurement theory to focus on designing measurement instruments that are both useful and used in practice, i.e., that are aligned with practitioner needs and designed to overcome barriers impeding their uptake~\cite{glasgow_pragmatic_2013}. Drawing on work from measurement theory and pragmatic measurement, we identify opportunities to improve measurement instruments~and~their~uptake among practitioners.\looseness=-1

\section{Related Work}\label{s-background}
The NLP research community has made publicly available numerous instruments for measuring representational harms caused by LLM-based systems. We provide examples of such instruments in Table~\ref{t-instruments},\footnote{Table~\ref{t-instruments} is not exhaustive, nor is it intended to suggest that we are specifically critiquing the instruments that are listed.} and point the reader to recent surveys by~\citet{sheng-etal-2021-societal},~\citet{dev-etal-2022-measures}, and~\citet{gallegos-etal-2024-bias} for more complete overviews.\looseness=-1

Assessments of existing measurement instruments have identified potential limitations to their usefulness. Prior work has shown that the concepts such instruments seek to measure are often poorly motivated, unclear, or not meaningfully measured by those instruments~\cite{blodgett-etal-2020-language, blodgett-etal-2021-stereotyping, goldfarb-tarrant-etal-2023-prompt, xiao-etal-2023-evaluating-evaluation, delobelle-etal-2024-metrics, porada-etal-2024-challenges, zhao_position_2024}. Researchers have also found that instruments can be highly sensitive to implementation choices that should not affect measurement outcomes~\cite{antoniak-mimno-2021-bad, delobelle-etal-2022-measuring, seshadri2022quantifying, sclar2023quantifying, shu-etal-2024-dont}. Furthermore, instruments often fail to produce measurements that enable informed actions~\cite{delobelle-etal-2024-metrics}, or that are useful for measuring representational harms in real-world deployment contexts~\cite{goldfarb-tarrant-etal-2021-intrinsic, cao-etal-2022-intrinsic, delobelle-etal-2022-measuring}.\looseness=-1

In this paper, we explore the extent to which these assessments of existing measurement instruments reflect the needs of practitioners, which are often both implicit and impacted by practical constraints~\cite{zhou-etal-2022-deconstructing}. We build on prior work from HCI showing that practitioners often struggle to use artifacts developed by researchers, in part because these artifacts are misaligned with practitioner needs~\cite{holstein_improving_2019, lee_landscape_2021, richardson_towards_2021, deng_exploring_2022, balayn__2023, ojewale_towards_2024}. This work focused on the needs of practitioners who seek to measure allocative harms caused by predictive models. To our knowledge, we are the first to explore whether and to what extent publicly available measurement instruments meet the real-world needs of practitioners who seek to measure representational harms caused by LLM-based systems.\looseness=-1

\vspace{1.85cm}
\section{Methods}\label{s-method}
To understand the extent to which publicly available measurement instruments meet the needs of practitioners, we conducted a series of semi-structured interviews with 12 practitioners tasked with evaluating LLM-based systems.  Participants, whom we refer to throughout by IDs P1--P12, worked on LLM-based systems (e.g., search engines, chatbots) and content moderation tools for such systems. See Table~\ref{t-participants} for participant details.\footnote{Participants were based in North America and Europe. They self-identified as white, Asian, and Middle Eastern/North African; and as men, women, and non-binary. To preserve~anonymity, we do not report individual demographics.}
\looseness=-1

\begin{table}
\centering
\begin{small}
\begin{tabular}{lll} 
\toprule
\textbf{PID} & 
\textbf{Role} & \textbf{Employer} \\
\midrule
P01 & Research engineer & Big tech company \\
P02 & Applied scientist & Big tech company \\
P03 & Research scientist & Big tech company \\
P04 & Research engineer & Big tech company \\
P05 & Consultant & AI startup \\
P06 & Scientist & Big tech company \\
P07 & Research scientist & AI startup \\
P08 & Data engineer & Big non-tech company \\
P09 & NLP specialist & Big non-tech company \\
P10 & Research scientist & AI startup \\
P11 & Researcher & AI nonprofit \\
P12 & Researcher & Big tech company \\
\bottomrule
\end{tabular}
\caption{Participant details.}
\label{t-participants}
\vspace{-0.3cm}
\end{small}
\end{table}

\paragraph{Recruitment.} We recruited participants through our professional networks, social media, cold emails, and snowball sampling~\cite{morgan2008snowball}. Each interview was one hour long and conducted between June and August 2024. All participants provided informed consent prior to their interviews. Each participant received a \$75 gift card. The study was approved by Microsoft's research IRB.
\looseness=-1

\paragraph{Interviews.} 
To scaffold the interviews, we identified a set of desiderata for measurement instruments: validity, reliability, specificity, extensibility, scalability, interpretability, and actionability (see Table~\ref{t-desiderata}).  These desiderata were identified based on our own experiences measuring representational harms caused by LLM-based systems and a systematic review of the NLP literature on  assessing~measurement instruments (see Appendix~\ref{a-lit-review}).\looseness=-1 

We began each interview by asking the participant to describe their role and the LLM-based systems they worked on. Next, we asked them to walk us through an example of how they measured representational harms, noting the publicly available measurement instruments they used or considered using. We then asked them to reflect on their experiences with those instruments, discussing any challenges they faced. We prompted them about whether they faced any challenges related to instruments failing to meet any of the desiderata listed in Table~\ref{t-desiderata} and also asked open-ended questions about any other challenges they faced. Our semi-structured interview guide is in Appendix~\ref{a-interview-guide}.\looseness=-1

We conducted interviews until saturation, i.e., until multiple consecutive interviews did not uncover any new challenges~\cite{small_2009_how}. Nevertheless, we note that our recruitment efforts yielded a low response rate, which likely resulted in a skewed participant pool (we discuss this in our limitations section). Therefore, we are careful not to over-generalize our findings: our interviews were specifically intended to answer the question, ``what challenges do practitioners face when trying to use publicly available measurement instruments?'' but not questions like, ``what is the prevalence~of~these~challenges among all practitioners?''\looseness=-1

\begin{table}[!t]
\begin{small}
\centering
\begin{tabular}{ 
p{1.6cm} 
p{5.3cm}}
\toprule

\textbf{Desideratum} & 
\textbf{Definition}
\tabularnewline
\midrule

Validity & 
Meaningfully measures what stakeholders think it measures
\tabularnewline
\midrule

Reliability & 
Results in similar measurements when used in similar ways, especially over time
\tabularnewline
\midrule

Specificity & 
Sufficiently specific to a system, its use cases, and its deployment contexts
\tabularnewline
\midrule

Extensibility & 
Can be adapted for different systems, use cases, and deployment contexts
\tabularnewline
\midrule

Scalability & 
Can scale to increasing workloads
\tabularnewline
\midrule

Interpretability & 
Produces measurements that can be understood by stakeholders
\tabularnewline
\midrule

Actionability & 
Produces measurements that can be acted upon by stakeholders
\tabularnewline

\bottomrule

\end{tabular}
\caption{Desiderata for measurement instruments. Measurement instruments that fail to meet these desiderata may be challenging to use. We used these desiderata to scaffold the interviews (see the guide in Appendix~\ref{a-interview-guide}).}
\label{t-desiderata}
\vspace{-0.3cm}
\end{small}
\end{table}

\paragraph{Thematic analysis.} We conducted a thematic analysis using an inductive--deductive coding approach~\cite{braun_using_2006, braun_reflecting_2019}. We coded the interview transcripts for challenges mentioned by participants. Initially, the set of challenges that we focused on corresponded to the desiderata listed in Table~\ref{t-desiderata} (i.e., challenges that arose because an instrument failed to meet a given desideratum). The first author coded each transcript and, at that time, identified additional challenges raised by participants that were not covered by our original set of codes. The first and second authors discussed these uncategorized challenges to create an expanded set of codes, and the first author reread and re-coded each transcript. All authors discussed and synthesized the codes into themes based on how, why, and to what extent the associated challenges impacted practitioners' abilities to use measurement instruments. Finally, at least one author other than the first author re-coded each transcript. Any coding disagreements identified through that process were resolved~through discussion with all of the authors.\looseness=-1

\paragraph{Positionality.}
We are a group of researchers and practitioners with expertise in the domains of NLP, machine learning, statistics, computational social science, software engineering, and responsible AI. Collectively, we are familiar with the NLP research community as well as the needs of practitioners working on NLP products and services. Many of us are employed in industry positions in which we have been tasked with measuring representational harms caused by LLM-based systems and have personally faced some of the same challenges raised by participants. Our professional experiences measuring representational harms had an impact on our research, in particular by influencing the set of desiderata that we used to scaffold the interviews.\looseness=-1
\section{Practitioner Challenges}\label{s-results}

Participants reported being aware of a range of publicly available measurement instruments, including tools designed to identify unsafe or toxic text (e.g., DeBERTa~\cite{he2021deberta} and Llama Guard~\cite{inan2023llamaguardllmbasedinputoutput}), as well as datasets and benchmarks (e.g., StereoSet~\cite{nadeem-etal-2021-stereoset} and CrowS-Pairs~\cite{nangia-etal-2020-crows}). However, these measurement instruments 
were used by at most one or two participants---none were more widely used. More importantly, all participants discussed facing challenges that prevented them from using publicly available measurement instruments.\looseness=-1

Specifically, participants experienced two types of challenges that left them unable to use publicly available measurement instruments, even when they had a desire to do so. First, instruments are \textit{not useful} when they do not meaningfully measure what practitioners seek to measure or are otherwise misaligned with practitioner needs (\S\ref{s-results-not-useful}). Second, instruments are \textit{not used} when practitioners face barriers impeding their uptake---even if the instruments are useful (\S\ref{s-results-not-used}). These challenges are context-dependent and determined by practitioner needs; the same instrument may be useful for one~practitioner in one context but not for another.\looseness=-1 

We found that the desiderata identified in Table~\ref{t-desiderata} aligned closely with considerations participants described as being central to their decisions about the usefulness of measurement instruments. Validity and specificity were primary considerations---all participants reported that they considered these desiderata and chose not to use measurement instruments that did not meet them. Participants also considered interpretability and actionability, but typically not until after they had already deemed measurement instruments sufficiently valid and specific. Although participants reported being concerned about the reliability and scalability of measurement instruments, most did not consider these desiderata when deciding whether to use an instrument. Participants did not report considering any additional desiderata.\footnote{We do not discuss extensibility here because participants did not report explicitly considering extensibility when deciding whether to use measurement instruments. Instead, we discuss it in \S\ref{s-bridging-gaps} as a desideratum that can be used to~improve~the usefulness and uptake of measurement instruments.\looseness=-1} However, participants identified additional challenges to using measurement instruments in the form of practical and institutional barriers impeding their uptake. Finally, participants also identified cases where challenges were exacerbated by the specifics of measuring representational harms or~evaluating LLM-based systems (\S\ref{s-results-exacerbated}).\looseness=-1

\subsection{Measurement Instruments Are Not Useful}\label{s-results-not-useful}
\begin{quote}
\textit{``Does it result in valid measurements?\ldots Is [it] going to translate well to my scenario?'' -- P2}
\end{quote}
Every participant reported at least one experience in which they chose not to use a measurement instrument due to concerns related to one or more desiderata. In fact, multiple participants reported that they were unable to use any publicly available measurement instruments because of challenges related to their usefulness (P1, P2, P6, P8, P9).\looseness=-1

\paragraph{Practitioners cannot use measurement instruments that they perceive as lacking \underline{validity}.}
Many participants reported feeling unable to use measurement instruments because the concepts that those instruments were intended to measure were not clearly defined or linked to any existing theoretical understandings of those concepts (P3, P6--P8, P10--P12). This makes it very difficult to pose the question of ``how well'' those concepts are being measured. Even when concepts were clearly defined, participants reported that instruments sometimes failed to meaningfully measure them (``There is a Jigsaw dataset that is supposed to be used to measure gender biases in hate speech detection systems. But when I [looked] into the data, the samples were often not about gender biases at all. They just [included gender] keywords\ldots I concluded that it's simply not up to the task.'' -- P11, also P3--P6, P12). As another example, participants reported frequently encountering datasets that contain mislabeled instances (``Every single public benchmark we use\ldots has a couple of rows that we look at by eye, and we're like, `that doesn't make sense.' And then that makes us question the entire validity of the benchmark.'' -- P7, also P3--P6, P11). These potential threats to validity caused participants to lose trust in measurement instruments.\looseness=-1

\paragraph{Data contamination exacerbates \underline{validity} concerns.}
Data contamination was an overarching concern about the validity of publicly available datasets and benchmarks. Because  developers of LLMs seldom disclose their training data, it is difficult to tell whether an LLM-based system that performs well on a benchmark has simply been trained using the benchmark data. Half of the participants expressed discomfort with using publicly available benchmarks and datasets under~any circumstances (P4--P6, P9, P10, P12).\looseness=-1

\paragraph{Practitioners cannot use instruments that lack \underline{specificity} for their needs.}
All participants reported choosing not to use publicly available measurement instruments because they were not sufficiently specific to their needs. Many reported creating their own instruments from proprietary data as a result (``we had to develop tests that were more suited to the sorts of scenarios that would happen in the workplace, not just a chat conversation that would happen outside of the workplace.'' -- P2, also P1, P3, P5, P6, P8--P12).\looseness=-1

\paragraph{The contextual nature of representational harms exacerbates \underline{specificity} concerns.}
Participants reported that many publicly available measurement instruments are not sufficiently specific to the representational harms they sought to measure. They attributed this to the fact that measuring representational harms requires cultural context; i.e., a specific understanding of who may be harmed and how (``it's tough to come in and say, here's this dataset with a bunch of stereotypes about race. Hopefully all of these stereotypes are gonna be present in this very specific system that we're working on.'' -- P3, also P5, P6, P8, P9, P11).\looseness=-1

\paragraph{Practitioners struggle to use measurement instruments that lack \underline{interpretability}.}
Several participants reported experiencing concerns about whether instruments produce measurements that can be understood (P1, P2, P5, P9, P11, P12). In particular, participants felt that measurements produced by tools and benchmarks could not be interpreted without additional information. As P11 put it, after using a measurement instrument, ``you end up with a number'' and then must decide ``when will this number become problematic?'' Without more information about what measurements mean (e.g., comparisons to other measurements from the same instrument), it~is~challenging to understand them.\looseness=-1

\paragraph{Practitioners deprioritize measurement instruments that they perceive as lacking \underline{actionability}.} 
Some participants reported experiencing concerns about whether instruments produce measurements that can be acted upon (P2, P3, P5--P7, P10). These concerns often stemmed from other issues, e.g., because measurement instruments are not sufficiently valid (P7) or interpretable (P10) for stakeholders to confidently act upon their outputs. 
Other participants reported that if they could not pre-identify a clear strategy for mitigating a harm, they often deprioritized measuring it (``You only have so many hours in a week\ldots the [harms] that don't have a clear mitigation strategy\ldots they're unlikely to be useful [to measure].'' -- P3, also P2, P5, P6, P10).\looseness=-1

\paragraph{While \underline{reliability} is desirable in theory, it is often not considered in practice.}
Participants did not report observing reliability issues in publicly available measurement instruments and then choosing not to use those instruments as a result---despite the importance of reliability to measurement~\cite{jacobs_measurement_2021} and the well-documented lack of reliability exhibited by measurement instruments in the NLP literature~\cite[e.g.][]{sclar2023quantifying, shu-etal-2024-dont, delobelle-etal-2024-metrics}. We hypothesize that this mismatch may be due to the fact that validity and specificity were primary considerations for the participants we interviewed, followed by interpretability and actionability. If instruments failed to meet these desiderata, participants chose not to use them---never reaching the point of considering whether the instruments were sufficiently reliable.\looseness=-1

\paragraph{\underline{Scalability} concerns do not typically prevent practitioners from using measurement instruments.}
Most commonly, participants reported scalability challenges related to measurement instruments that required repeated calls to LLM-based systems (e.g., instruments that rely on LLMs as judges, or instruments that require multiple responses from LLM-based systems to produce measurements) (P1--P3, P11). Although participants noted that repeated calls to LLM-based systems added time, costs, and environmental impacts to their measurement processes, they considered this an inevitable result of evaluating LLM-based systems. However, when participants sought to do online measurement for client-facing systems, they sometimes chose not to use such instruments due to the latency or token limits imposed by those instruments (P7, P10). We hypothesize that when practitioners exclusively do offline measurement, scalability of the instruments themselves is less of a concern; however, when online measurement is the goal, scalability concerns become more salient.
\looseness=-1

\subsection{Measurement Instruments Are Not Used}
\label{s-results-not-used}
\begin{quote}
    \textit{``[I]t can just be really hard to actually get the time allotted to go and find the resources that probably exist. So usually we end up making our own thing.'' -- P3}
\end{quote}
Even when participants thought that publicly available measurement instruments might be useful, they were sometimes unable to use them due to practical and institutional barriers. These barriers arose not because the instruments failed to meet particular desiderata, but because of the challenges specific to measuring representational harms in practice. We highlight these barriers not to suggest that designers of measurement instruments are solely responsible for addressing them, but because it is important to identify barriers impeding the uptake of measurement instruments to provide a more holistic view of what can~be changed and who can make those changes.\looseness=-1

\paragraph{Practitioners face \underline{practical barriers} to using measurement instruments when those instruments do not meet organizational requirements.} 
Participants reported being unable to use publicly available instruments due to challenges related to organizational requirements, such as 
security (``We have this agreement not to disclose [our customers' data] outside. With these publicly available tools, especially with the ones that are not open source, we~have always these security~issues.''~--~P9) and data licensing (``We have to use compliant datasets. In some case, we just created our own [version] of a dataset that we cannot use.'' -- P4, also P6).\looseness=-1

\paragraph{Practitioners face \underline{institutional barriers} when organizational culture impedes uptake of measurement instruments.} Participants reported facing a lack of organizational support---and sometimes outright disincentives---for using publicly available instruments, regardless of whether those instruments might be useful. For example, participants reported that they sometimes created new measurement instruments due to a lack of time to find existing publicly available instruments (P3, P7). Some participants were not able to use measurement instruments if those instruments did not align with their organizations' processes. For example, P10 reported that their organization had a preference for measurement processes that were similar to those used in software engineering, i.e., having a small set of curated test cases that a system must pass prior to deployment. It can be unclear how to translate large benchmarks into smaller sets of prioritized test cases, e.g., ``the 30 cases that should not fail if we change anything'' (P10). Finally, participants reported facing limited incentives to measure representational harms, especially compared to other kinds of harms (e.g.,~quality-of-service harms) (P5, P7, P8, P10).\looseness=-1

\subsection{Challenges Are Exacerbated by, but Extend Beyond, Representational Harms}\label{s-results-exacerbated}
Participants identified properties of representational harms that made them particularly challenging to measure. These included the fact that measuring representational harms requires additional information or expertise, such as cultural context or social science expertise (P2, P11). Additionally, some participants felt that the contestedness of representational harms made it more difficult to assess the validity or reliability of instruments intended to measure them (P4, P5, P9, P10, P12).\looseness=-1

Although our interviews focused specifically on instruments for measuring representational harms, participants consistently voiced that their challenges applied broadly to instruments for measuring other abstract or contested concepts.\footnote{These challenges did not apply to instruments intended to measure directly observable concepts, such as whether~an~LLM-based system~generates phone numbers (P7).\looseness=-1}  For example, participants also reported validity concerns about instruments for measuring disinformation (P11); specificity concerns about instruments for measuring the legality of text (because legal codes are specific to geographic jurisdictions) (P10); and interpretability and actionability concerns~about~machine translation benchmarks (P5).\looseness=-1

Finally, we note that multiple participants lived in countries where English is not the primary language, but nevertheless focused on evaluating LLM-based systems in English. Although we did not specifically recruit practitioners who were multilingual or focused on low-resource languages, a third of the participants discussed speaking languages other than English and challenges related to measuring representational harms in those languages (P5, P9, P11, P12). P11, who speaks a low-resource European language, shared that they had considered trying to develop a measurement instrument in that language, but opted against it due to the low probability that their instrument would be widely adopted. We therefore hypothesize that practitioners who seek to measure representational harms in low-resource languages likely face additional challenges related to the availability of~measurement instruments in those languages.\looseness=-1
\section{Addressing Practitioner Challenges}\label{s-bridging-gaps}
In this section, we draw on measurement theory from the social sciences and pragmatic measurement to improve the usefulness and uptake of measurement instruments. As we mentioned in \S\ref{sec:introduction}, measurement theory has long been concerned with designing instruments that are useful, while pragmatic measurement builds on measurement theory to focus on designing measurement instruments that are both useful and used in practice. Designers of measurement instruments can draw on measurement theory to improve the validity and reliability of their instruments, and can draw on pragmatic measurement to improve other aspects of the usefulness and uptake of their instruments. This is not to suggest that designers of measurement instruments are solely responsible for meeting practitioner needs. Rather, practitioners who seek to use measurement instruments are responsible for adapting those instruments to meet their specific needs. However, designers can facilitate this by designing extensible measurement instruments, which can help address usefulness and uptake challenges, as we explain below. Regulators and organizations that develop and deploy LLM-based systems can also play a role in removing both practical and institutional barriers. Finally, we briefly discuss  trade-offs inherent to attempting~to~meet~multiple desiderata simultaneously.\looseness=-1

\paragraph{Measurement theory provides a framework with which to improve the usefulness of measurement instruments.}
All participants reported concerns about whether measurement instruments meaningfully measure the concepts that they are intended to measure---often because those concepts were not clearly defined. These concerns were exacerbated by the fact that representational harms are abstract and contested, meaning that different measurement instruments may operationalize different definitions of representational harms. Measurement theory offers a framework with which these concerns~can~be better understood and addressed.\looseness=-1

Measurement theory provides a framework for obtaining measurements (e.g., scores calculated using a benchmark) of abstract concepts (e.g., ``stereotypes'') through the processes of systematization, operationalization, application, and interrogation~\cite{adcock_measurement_2001, wallach2025positionevaluatinggenerativeai}.  \textit{Systematization} is the process of formulating a specific, often theoretically grounded, definition of the concept of interest. This definition---the \textit{systematized concept}---is, as noted by multiple participants and in prior work, often absent from instruments for measuring representational harms.  Instead, designers of measurement instruments often jump straight to \textit{operationalization}, which is the process of developing one or more instruments for measuring~the~concept of interest.\footnote{We note that \textit{application} is the process of using the resulting measurement instruments to obtain measurements of the concept of interest, while \textit{interrogation} is the process of interrogating the validity of the systematized concept, the measurement instruments, and their resulting measurements.\looseness=-1}\looseness=-1

\paragraph{Designers of measurement instruments should draw on measurement theory to ensure that their instruments are \underline{valid} and \underline{reliable}.}
Jumping straight to operationalization not only creates uncertainty concerning what precisely measurement instruments are intended to measure, but also renders it impossible to pose the question of ``how well'' a concept of interest is being measured when that concept has multiple competing meanings. For example, stereotypes can be negative, neutral, or positive in sentiment.  An instrument that performs well at measuring stereotypes with negative sentiment may perform poorly if its validity is interrogated with respect to a broader definition of stereotyping that encompasses other sentiments. We therefore recommend that designers of measurement instruments do not skip systematization (or conflate it with operationalization) and clearly document systematized concepts as part of making~measurement instruments publicly available.\looseness=-1

Once a concept has been systematized, questions of how accurately the systematized concept is being measured can be answered using different lenses of validity and reliability~\citep{jacobs_measurement_2021}.  Because participants reported perceived threats to the validity of measurement instruments, we recommend that designers of measurement instruments use these lenses to rigorously interrogate the validity and reliability of their instruments, providing evidence from these interrogations when making their instruments publicly available~\cite[see][]{xiao-etal-2023-evaluating-evaluation, van_der_wal_undesirable_2024}. Lastly, we recommend that designers of measurement instruments release resources (e.g., guidelines, code) that practitioners can use to interrogate~validity and reliability in different contexts.\looseness=-1

\paragraph{Pragmatic measurement suggests additional ways to improve the usefulness of measurement instruments.} 
All participants struggled to use measurement instruments that were misaligned with their needs. We argue that drawing on pragmatic measurement, which builds on measurement theory and has historically been concerned with improving the real-world effectiveness of clinical research, can help designers of measurement instruments to improve aspects of the usefulness of their instruments beyond validity~and~reliability~\cite{glasgow_pragmatic_2013}.\looseness=-1

\paragraph{Designers of measurement instruments should draw on pragmatic measurement to ensure that their instruments are \underline{interpretable}, \underline{actionable}, and \underline{scalable} in different contexts.}
Pragmatic measurement suggests ways to improve the interpretability, actionability, and scalability of measurement instruments, all of which contribute to their usefulness. Many participants reported struggling to use measurement instruments that lack interpretabilty. Pragmatic measurement offers several ways to improve the interpretability of measurement instruments. For example, where possible, designers of measurement instruments should publish distributions of measurements produced by their instruments for known datasets~\cite{lewis_psychometric_2021}.\footnote{This is different from the practice of publishing the accuracy of measurement instruments on known datasets, which does~not~help practitioners interpret individual measurements.\looseness=-1} Additionally, designers of measurement instruments should publish information about how to interpret the measurements produced by their instruments (e.g., what does a measurement of 0.3 mean vs. a measurement of 0.8?)~\cite{stanick_pragmatic_2021}. These suggestions are simple, evidence-based ways to improve interpretability. Furthermore, they are complementary to the recommendations for improving actionability recently proposed by~\citet{delobelle-etal-2024-metrics}. See \citet{estabrooks_harmonized_2012},~\citet{martinez_instrumentation_2014}, and~\citet{albers_advancing_2020} for other ways to further improve~the~usefulness of measurement instruments.\looseness=-1

\paragraph{Practitioners are responsible for adapting measurement instruments to meet their \underline{specific} needs---but designers of measurement instruments can help by ensuring that their instruments are \underline{extensible}.}
Pragmatic measurement emphasizes the importance of \textit{extensibility}, i.e., whether an instrument can be adapted for different systems, use cases, and deployment contexts.\footnote{Extensibility is sometimes also called \textit{adaptability}.\looseness=-1} In pragmatic measurement, extensibility is a low-effort, high-impact way to make measurement instruments useful: rather than designing multiple measurement instruments to meet different needs, designers can focus on extensible instruments that practitioners can individually tailor~\cite{powell_refined_2015, waltz_use_2015}. Based on our findings in \S\ref{s-results} and the pragmatic measurement literature, we identify key criteria that need to be met for measurement instruments to be considered extensible. First, we recommend making them open source or otherwise modifiable. Second, we recommend that they be \textit{modular}, i.e., composed of discrete, interconnected pieces~\cite{baldwin_design_2000}.\looseness=-1

\paragraph{Designers of measurement instruments and practitioners should both ensure that measurement instruments remain \underline{valid} and \underline{reliable} when adapted to new contexts.}
A measurement instrument may be sufficiently valid and reliable in one context, but not in another. To ensure that extensible measurement instruments remain valid and reliable, we recommend that designers of measurement instruments be explicit about what is being measured (the systematized concept) and how it is being measured. Separating systematization from operationalization can enable measurement instruments to be more easily modified. Designers of measurement instruments should also be explicit about the impacts on validity and reliability of modifying different aspects of their instruments when adapting them to new contexts~\cite{powell_refined_2015, waltz_use_2015}. We also emphasize that any time a measurement instrument is to be used in a new context, its validity and reliability must be re-interrogated~\cite{martinez_instrumentation_2014}. To facilitate this, we reiterate that designers of measurement instruments should release resources (e.g., guidelines, code) that practitioners can use to interrogate validity and reliability in different contexts. Designers should also make it easy for practitioners to compare measurements produced by instruments with and without adaptations~\cite{martinez_instrumentation_2014}.\looseness=-1

\paragraph{Extensibility can help overcome \underline{practical barriers} impeding the uptake of measurement instruments.} Participants noted practical barriers to using measurement instruments, including security and data licensing. Extensible measurement instruments that can adapted to support local use may help mitigate some security concerns. It should also be possible to adapt extensible measurement instruments to incorporate newly obtained, proprietary, or other sources of~data~in order to overcome data licensing issues.\looseness=-1

\paragraph{Trade-offs.}
Designing measurement instruments that are simultaneously valid, reliable, specific, extensible, scalable, interpretable, and actionable and are actually used by practitioners is a tall order. Trade-offs are inevitable: for example, as measurement instruments become more specific to a particular context, they are likely to become less extensible. We emphasize the importance of interrogating validity and reliability, and, beyond that, we caution against over-indexing on any particular desideratum. For example, designers of measurement instruments who optimize for scalability without considering other desiderata may face unwanted trade-offs. By analogy, blocklists, which are commonly used to mitigate representational harms, are relatively low effort to deploy and scale, but do not wholly and fully capture the contextual nature of representational harms. As a result, they produce false positives when social groups engage in actions like reclaiming slurs, which can lead to over-moderation or erasure of those groups~\cite{vashishtha-etal-2023-performance}.\looseness=-1

\section{Conclusion}
We found that practitioners are often unable to use publicly available instruments for measuring representational harms and identified two types of challenges. In some cases, instruments are \textit{not useful} because they do not meaningfully measure what practitioners seek to measure or are otherwise misaligned with practitioner needs. In other cases, instruments---even useful instruments---are \textit{not used} by practitioners due to practical and institutional barriers impeding their uptake. Furthermore, both types of challenges can be exacerbated by the specifics of measuring representational harms or evaluating LLM-based systems. Drawing on measurement theory and pragmatic measurement, we provided recommendations for addressing these~challenges to better meet practitioner needs.\looseness=-1
\section*{Limitations}\label{s-limitations}
The primary limitation of our paper is that our recruitment efforts yielded a low response rate. As is the case with many studies targeting technology workers~\cite{scheuerman_walled_2024}, it was challenging to identify and recruit potential participants. For example, in addition to relying on our professional networks and social media, we cold-emailed 73 practitioners whom we identified as being potential participants---based on LinkedIn profiles, company websites, and publications at ACL venues---and whose contact information was available online. We were ultimately able to interview one such practitioner. Potential participants often declined to speak with us due to NDAs or other confidentiality concerns. We interviewed a total of 12 practitioners, some of whom declined to answer certain questions in order to remain in compliance with their organizations' NDAs. Due to these recruitment challenges, it is likely that our participant pool is not broadly representative of all practitioners. For example, it is likely skewed toward practitioners who faced challenges when trying to use publicly available instruments for measuring representational harms caused by LLM-based systems. Therefore, we are careful not to over-generalize our findings. For example, our findings do not enable us to answer questions about the prevalence of the challenges that we identified. We believe there is potential to expand on our findings in future work by surveying practitioners to answer such questions.\looseness=-1

Because qualitative research of this type is not typical in ACL venues, we clarify that the above limitation is not the same as having a too-small sample size. Our sample size is, in fact, typical of HCI interview studies~\cite{caine_local_2016}. Indeed, we conducted interviews until saturation, i.e., until multiple consecutive interviews did not uncover any new challenges~\cite{small_2009_how, hennink_sample_2022}. We emphasize that in interview studies like ours, the goal is not to interview a target number of participants. Rather, the process of establishing saturation determines the sample size~and whether it is appropriate~\cite[see][]{small_2009_how}.\looseness=-1

Finally, because English is the only language shared by all members of the research team, we were only able to conduct interviews in English. Our findings are therefore centered on practitioners who focused (either primarily or exclusively) on evaluating LLM-based systems in English, although some participants did touch on low-resource languages in \S\ref{s-results}. Nevertheless, we hypothesize that practitioners who seek to measure representational harms in low-resource languages likely face additional challenges~\cite{benderrule}.\looseness=-1

\section*{Ethical Considerations}

We do not anticipate any risks to society or the general public associated with our findings as we simply identified practitioners' challenges to using publicly available instruments for measuring representational harms caused by LLM-based systems.\looseness=-1

Our research involved interviewing humans.\footnote{Our semi-structured interview guide is in Appendix \ref{a-interview-guide}, and we provide details about participant recruitment in \S\ref{s-method}.~The study was approved by Microsoft's research IRB.\looseness=-1} As a result, there is an inherent risk to our participants: the possibility of identification, as we did collect personally identifiable information (in order to obtain informed consent and record interviews). We have taken the following steps to reduce this risk. First, we created de-identified interview transcripts and then deleted the original recordings. We saved identified consent forms separately from the de-identified interview transcripts (prior to publication, we maintained a separate file linking participants' names and IDs to ensure that if a participant contacted us asking for their data to be removed, we would be able to do so). We also asked participants not to share confidential information with us during the interview process. Additionally, we allowed all participants to review direct quotes so they could request that we remove any information~that might enable them to be identified.\looseness=-1

\section*{Acknowledgments}
We thank members of Microsoft Research's Sociotechnical Alignment Center (Chad Atalla, Emily Corvi, Alex Dow, Nick Pangakis, Stefanie Reed, Dan Vann, Matt Vogel, and Hannah Washington) for their invaluable feedback and participation in our pilot studies. We also thank members of Microsoft Research's FATE group. Finally, we~thank the practitioners who agreed to be interviewed.\looseness=-1

\bibliography{anthology,custom}

\clearpage
\appendix

\section{Systematic Literature Review to Identify Desiderata}\label{a-lit-review}

\begin{figure*}[t]
\centering
{\includegraphics[width=\textwidth]{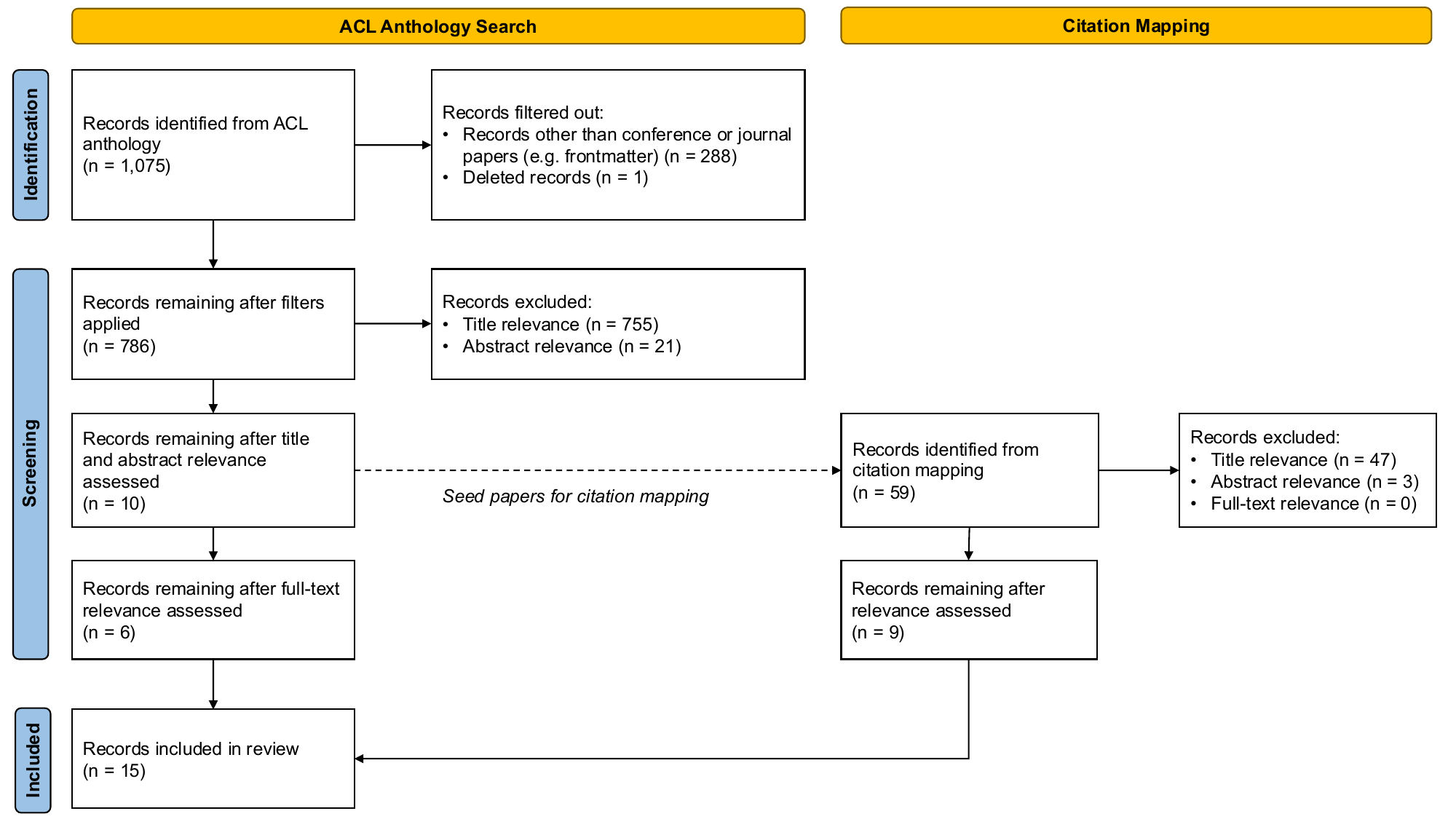}}
\caption{Our PRISMA flow diagram.
\label{f-prisma}}
\end{figure*}

To identify the set of desiderata that we used to scaffold our interviews, we conducted a systematic review of the NLP literature on assessing measurement instruments. To do this, we followed PRISMA guidelines~\cite{PRISMA}. We~provide our PRISMA flow diagram in Figure~\ref{f-prisma}.\looseness=-1

\subsection{Search Strategy}
\paragraph{ACL Anthology Search.} Using the ACL Anthology API, we identified all papers published in ACL venues through 2024 that met the following criteria:\looseness=-1
\begin{itemize}
    \item Title contains: `eval*' OR `measur*' 
    \item Abstract contains: `meta' OR `survey*' OR `review*' OR `assess*' OR `audit*'
\end{itemize}
This search produced 1,075 results. We reviewed these search results for title and abstract relevance (see \S\ref{app:selection_strategy}) in order to identify ten papers that were about assessing measurement instruments. We then used these ten papers as ``seed papers'' to~conduct~an additional~citation mapping search.\looseness=-1

\paragraph{Citation Mapping Search.} To ensure that we captured relevant work that was not published in ACL venues or that may have been missed by our keyword search, we used the Semantic Scholar API to identify all papers that were referenced by or cited the ten seed papers. We then selected all papers that were referenced by or cited at least two of the seed papers, resulting in an additional 59 papers, which we reviewed for title and abstract relevance.\looseness=-1

\subsubsection{Selection Strategy}
\label{app:selection_strategy}
To screen the results of our ACL Anthology keyword search, we focused on only those papers that were published in ACL conferences or journals (i.e., not front matter, tutorials, or workshop papers) and that had not been marked as deleted. After screening the papers, the first author iteratively reviewed each remaining paper for title relevance and,  if the title was deemed relevant, abstract relevance. Titles and abstracts were considered relevant if they indicated that the papers appeared to be about 1) assessing measurement instruments, 2) instruments for measuring  abstract or contested concepts from text (e.g., not about information retrieval, not about visual or audio tasks), and 3) measurement instruments (e.g., not solely about annotator reliability or researchers' misuse of measurement instruments). We used the same title and abstract relevance criteria to screen the results of the citation mapping search. Finally, the first author evaluated all remaining papers for full-text relevance, again considering papers relevant if they were about the topics listed above.\looseness=-1

\subsubsection{Annotation Strategy}
The first author read each paper and extracted all passages of text that appeared to identify a particular desideratum of measurement instruments (e.g., text identifying a quality that measurement instruments should have, or text identifying a quality such that instruments lacking that quality are challenging to use). All authors other than the first author developed an initial list of desiderata (validity, reliability, specificity, extensibility, scalability, interpretability, and actionability; see Table~\ref{t-desiderata}) based on their experiences measuring representational harms caused by LLM-based systems. In discussion with the other authors, the first author mapped each passage to these desiderata. While conducting this exercise, we did not identify any additional desiderata.  In other words, we were able to map all extracted passages~to~one of the desiderata listed in Table~\ref{t-desiderata}.
\looseness=-1

\subsubsection{Results}
We list the papers mentioning each desideratum below. Some papers mention the desideratum explicitly. Others mention the desideratum implicitly (e.g., not using the exact terminology we do, but describing the concept captured by the desideratum).\looseness=-1

\paragraph{Mentioned validity.} \citet{blodgett-etal-2020-language, blodgett-etal-2021-stereotyping, delobelle-etal-2022-measuring, delobelle-etal-2024-metrics, du-etal-2021-assessing, gehrmann_repairing_2023, goldfarb-tarrant-etal-2021-intrinsic, goldfarb-tarrant-etal-2023-prompt, novikova-etal-2017-need, reiter-2018-structured, sun-etal-2023-validity, van_der_wal_undesirable_2024, xiao-etal-2023-evaluating-evaluation, zhou-etal-2022-deconstructing}.

\paragraph{Mentioned reliability.} \citet{blodgett-etal-2021-stereotyping, delobelle-etal-2022-measuring, delobelle-etal-2024-metrics, du-etal-2021-assessing, gehrmann_repairing_2023, goldfarb-tarrant-etal-2023-prompt, novikova-etal-2017-need, seshadri2022quantifying, sun-etal-2023-validity, van_der_wal_undesirable_2024, xiao-etal-2023-evaluating-evaluation, zhou-etal-2022-deconstructing}.

\paragraph{Mentioned specificity.}
\citet{delobelle-etal-2024-metrics, du-etal-2021-assessing, gehrmann_repairing_2023, van_der_wal_undesirable_2024, zhou-etal-2022-deconstructing}.

\paragraph{Mentioned extensibility.}
\citet{gehrmann_repairing_2023, reiter-2018-structured, zhou-etal-2022-deconstructing}.

\paragraph{Mentioned scalability.}
\citet{gehrmann_repairing_2023, novikova-etal-2017-need, xiao-etal-2023-evaluating-evaluation, zhou-etal-2022-deconstructing}.

\paragraph{Mentioned interpretability.}
\citet{delobelle-etal-2024-metrics, du-etal-2021-assessing, gehrmann_repairing_2023, van_der_wal_undesirable_2024, xiao-etal-2023-evaluating-evaluation}.

\paragraph{Mentioned actionability.}
\citet{delobelle-etal-2024-metrics}.

\section{Semi-Structured Interview Guide}\label{a-interview-guide}
Our semi-structured interview guide is shown below. As is typical of semi-structured interviews, not every participant was asked exactly the same questions in exactly the same order, and some participants were asked additional follow-up or clarifying questions based on the answers they provided. The interview questions were supplemented with a set of slides containing definitions of key terms that we screenshared with participants. The~definitions~are included in the script below.\looseness=-1

\subsection{Introductions [5 min]}
Welcome! Thank you so much for taking the time for this interview. Before we get started, I just want to quickly introduce myself, talk about the goals of this study, and give you a chance to ask any questions you might have. This research study is intended to understand gaps between research and practice in evaluating large language model (LLM)-based systems, with a focus on measuring harms, adverse impacts, or other undesirable behaviors. In this interview, I'll ask you to share your experiences with and opinions on such evaluations, without discussing confidential information. I will also record this interview for the purpose of creating a deidentified transcript. If you prefer that your video not be recorded, please feel free to turn your camera off at this time. In addition, if at any point you would like to skip a question, take a break,~or~end~the interview, please feel free to do so.\looseness=-1

Do you have any questions before we get started?

\subsection{Background [5 min]}
    \begin{itemize}[left=20pt]
        \item[\textbf{[Q1]}] To start, please briefly describe your role, focusing on your professional experience as it relates to LLM-based systems. 
        \item[\textbf{[Q2]}] Can you briefly describe the LLM-based system(s) that you have previously evaluated, currently evaluate, or plan to evaluate? 
    \end{itemize}
\subsection{Experience with measurement instruments for representational harms [15 min]}
    \begin{itemize}[left=20pt]
        \item[\textbf{[Q3]}] Throughout this interview, I will be focusing primarily on representational harms, which occur when ``a system represents some social groups in a less favorable light than it represents other groups by stereotyping them, demeaning them, or failing to recognize their existence altogether.'' 
        \item[] What examples of representational harms caused by LLM-based systems are you aware of? 
        \item[] \textit{If interviewee was not familiar with representational harms, we provided the following examples:}
        \begin{itemize}[left=0pt]
        \item[--] LLMs might reinforce stereotypes, for example, by using the word ``nurse'' to refer to a female healthcare provider and the word ``doctor'' to refer to a male healthcare provider in otherwise identical contexts.
        \item[--] LLMs might generate slurs or derogatory language about a social group.
        \item[--] LLMs might erase a social group, for example, by only listing male athletes when a user asks for examples of talented soccer players, thus failing to recognize the existence of non-male soccer players.
        \end{itemize}
        \item[\textbf{[Q4]}] Do your previous, current, or planned evaluation(s) of LLM-based system(s) involve measuring representational harms? 
    \item[\textbf{[Q5]}] What types of representational harms are you measuring? 
    \item[\textbf{[Q6]}] Can you walk me through, from start to finish, an example of how you measure representational harms? I'm especially interested in hearing about how you decided on your approach, whether you relied on existing, publicly available tools, benchmarks, datasets, metrics, annotation guidelines, and so on, or whether you decided to develop your own. 
    \item[] \textit{To allow for open-ended discussion, we did not provide participants with a specific definition of `measurement instruments'; rather, we provided the following examples of instruments:}
    \begin{itemize}[left=0pt]
    \item[--] An example of a tool is Perspective API.
    \item[--] An example of a benchmark is StereoSet, which includes a dataset of prompts that could elicit stereotyping content with corresponding metrics that measure the extent to which a language model produces stereotypes.
    \item[--] An example of a dataset is WildChat, which is a corpus of 1 million real user-ChatGPT interactions.
    \item[--] Examples of metrics are the Word and Sentence Embedding Association Tests (WEAT and SEAT), which measure whether “attribute words” (e.g. male, female) are disproportionately associated with a set of “target words” (e.g. different professions).
    \item[--] Annotation instructions are sets of instructions and examples for humans to use when annotating system outputs for particular properties.
    \item[--] An example of another type of instrument is Matched Guide Probing, a method adapted from sociolinguistics.
    \end{itemize}
\end{itemize}
\textit{For each instrument mentioned, we asked the following questions:}
\begin{itemize}[left=20pt]
        \item[\textbf{[Q7]}] What type(s) of representational harms are you measuring with [this instrument]? 
        \item[\textbf{[Q8]}] How did you decide to use [this instrument]? 
        \item[\textbf{[Q9]}] How do you use [this instrument] in your evaluation(s)? 
        \item[\textbf{[Q10]}] Where did [this instrument] come from? Did you develop it yourself, modify an existing [instrument], or use an existing [instrument] as-is? 
        \end{itemize}
\textit{If applicable, for one instrument that the interviewee developed themselves, we asked the following questions:}
    \begin{itemize}[left=20pt]
        \item[\textbf{[Q11]}] Why did you decide to develop [this instrument] yourself?  
        \item[\textbf{[Q12]}] What, if any, actions have you taken or plan to take upon seeing the measurements obtained using [this instrument]? 
    \end{itemize}
\textit{If applicable, for one instrument that the interviewee adapted from an existing instrument, we asked the following questions:}
    \begin{itemize}[left=20pt]
        \item[\textbf{[Q13]}] Why did you decide to start with this existing [instrument]?  
        \item[\textbf{[Q14]}] Why did you decide to modify [this instrument] rather than using it as-is? 
        \item[\textbf{[Q15]}] What, if any, actions have you taken or plan to take upon seeing the measurements obtained using [this instrument]? 
    \end{itemize}
\textit{If applicable, for one instrument that the interviewee used as-is, we asked the following questions:}
    \begin{itemize}[left=20pt]
        \item[\textbf{[Q16]}] Why did you decide to use this existing [instrument] as-is?  
        \item[\textbf{[Q17]}] What, if any, actions have you taken or plan to take upon seeing the measurements obtained using [this instrument]? 
    \end{itemize}
\subsection{Challenges with measurement instruments for representational harms [15 min]}
    \begin{itemize}[left=20pt]
    \item[\textbf{[Q18]}] Were there any other existing, publicly available [instruments] that you investigated using instead? 
        \item[\textbf{[Q19]}] \textit{For each instrument mentioned:} Why did you decide not to use [this instrument]? 
    \item[\textbf{[Q20]}] \textit{For each of the challenges defined below, say either:} 
    \item[] ``It sounds like you mentioned an issue to do with [challenge]. Is that correct?”, \textit{or} 
    \item[] “I don’t think you mentioned [challenge]. Did you experience any issues with this?'' 
    \item[] \textit{We provided interviewees with the following set of challenges related to measurement instruments:}
    \begin{itemize}[left=0pt]
    \item[--] Whether it results in valid measurements – i.e., meaningfully measures what stakeholders think it measures 
    \item[--] Whether it results in similar measurements when used in similar ways, especially over time 
    \item[--] Whether it is sufficiently specific to the system being evaluated and its particular use cases and deployment contexts
    \item[--] Whether it can scale to increasing workloads
    \item[--] Whether it can be adapted for different systems, use cases, and deployment contexts
    \item[--] Whether its resulting measurements can be understood by stakeholders
    \item[--] Whether its resulting measurements can be acted upon by stakeholders
    \end{itemize}
        \item[\textbf{[Q21]}] \textit{For each challenge experienced:}  What, if anything, did you do to address this issue? 
    \item[\textbf{[Q22]}] Did you experience any other issues that we haven’t discussed?  
        \item[\textbf{[Q23]}] \textit{If applicable:} What, if anything, did you do to address this issue? 
    \end{itemize}

\subsection{Comparing measurement of representational harms to other harms [5 min]}
\begin{itemize}[left=20pt]
    \item[\textbf{[Q24]}] Do your previous, current, or planned evaluation(s) of LLM-based system(s) involve measuring harms, adverse impacts, or other undesirable behaviors other than representational harms?  
    \item[\textbf{[Q25]}] \textit{If yes:} What types of harms, adverse impacts, or other undesirable behaviors?  
    \item[\textbf{[Q26]}] \textit{If yes:} Are your experiences measuring these types of harms, adverse impacts, or other undesirable behaviors similar to your experiences measuring representational harms? What, if anything, is similar and what, if anything, is different about your experiences? I'm especially interested in hearing about how the [instruments] you use to measure these types of harms, adverse impacts, or other undesirable behaviors are similar to or different from the [instruments] you use to measure representational harms. 
    \end{itemize}

\subsection{Desired improvements to measuring representational harms [5 min]} 
\begin{itemize}[left=20pt]
    \item[\textbf{[Q27]}] Putting aside any time or budget constraints, what, if anything, would you improve about the way that you previously, currently, or plan to measure representational harms?   
    \item[\textbf{[Q28]}] What do you need, that you don't currently have, in order to make those improvements? 
\end{itemize}

\subsection{Closing [5 min]}
\begin{itemize}[left=20pt]
    \item[\textbf{[Q29]}] Is there anything else you would like to tell us about your previous, current, or planned evaluation(s) of LLM-based system(s)?
\end{itemize}

\end{document}